 \documentstyle[twoside,fleqn,espcrc2]{article}

\newcommand{\AmS}{{\protect\the\textfont2
  A\kern-.1667em\lower.5ex\hbox{M}\kern-.125emS}}

\def\be{\begin{equation}}
\def\ee{\end{equation}}

\title{Massive fields in AdS(3) and compactification
in AdS spacetime\thanks{Based on talk given
at the International Conference ``Supersymmetry and Quantum Field
Theory'' (Kharkov, July 25-29, 2000)}
}

\author{R.R. Metsaev\address[MCSD]{
Department of Physics, The Ohio State University,\\
174 West 18th Avenue, Columbus, OH 43210-1106, USA}
        \thanks{Permanent address:
Department of Theoretical Physics, P.N. Lebedev Physical
Institute, Leninsky prospect 53,  Moscow 117924, Russia.
E-mail: metsaev@lpi.ru.
This  work  was  supported  by
the DOE grant DE-FG02-91ER-40690, by the INTAS project 991590,
by the RFBR Grant No.99-02-16207,
and RFBR Grant for Leading Scientific Schools, 
Grant No. 00-15-96566.}}

\begin{document}

\begin{abstract}
Massive arbitrary spin fields in AdS(3) spacetime are discussed in a
framework of light-cone gauge formulation. 
We also consider compactification  of AdS spacetime on manifold which is
warped product of AdS space and sphere.
Mass spectra of massless and massive fields upon 
such compactification are found.
\vspace{1pc}
\end{abstract}

\maketitle

\section{Introduction}

Motivated by a desire to understand stringy version of AdS/CFT 
correspondence a light-cone gauge formulation 
of string in $AdS_5 \times S^5$ 
Ramond Ramond background was recently developed\cite{1,2}. 
Brief review of this theme may be found in \cite{3}.
Discussion of alternative gauges for $AdS$ string may be found in 
\cite{4,5,6,7,1}. 
Various aspects of  $kappa$-invariant action of AdS string were 
investigated in \cite{8,9,10,11}.
In contrast to flat string the light-cone action of AdS string \cite{2}
is not quadratical in physical string coordinates - it is still to be
nonlinear in bosonic coordinates and involves terms of fourth degree in
fermionic coordinates. 
Understanding in which terms AdS string dynamics should be formulated 
is still to be challenging problem.
Anyway, since AdS superstring with RR charge
is available only in the light-cone gauge 
it seems that from the stringy perspective of
AdS/CFT correspondence the light-cone approach
is the only fruitful direction to go. 
Alternative approach which might open up new interesting possibilities
is based on twistor like variables \cite{101,102,103}.

In this work we develop light-cone gauge formulation
of massive arbitrary spin fields in $AdS_3$. This formulation may be
useful for the various reasons. On one hand it might be helpful
to understand string dynamics in $AdS_3$. 
Note that massless modes of $AdS_3$ string have already been described
in \cite{12} where the light-cone formulation 
of string in $AdS_3\times S^3$ background was developed.
So, our results may be relevant for discussion of
massive modes of $AdS_3$ string.
On the other hand by now due to \cite{13} 
consistent equations for interacting 
massless higher spin field theory in $AdS_4$ are known.
Because upon compactification on $AdS_3$ 
all massless higher spin modes turn out to be massive
one can expect that 
equations suggested in \cite{13} upon compactification
on $AdS_3$ will describe selfconsistent dynamics 
of massive arbitrary spin fields there.
It is not unlikely then that this compactified theory is related
with $AdS_3$ string theory or some subsector of the latter theory.
Another application of compactification we have in mind is a derivation
of dynamics of $AdS$ massive arbitrary spin field from the covariant
formulations of massless fields given in \cite{lop,14}.
To our knowledge compactification in AdS spacetime 
has not been discussed previously in the literature
and this is the second theme we study in this work.

\vskip-19cm
{} \hfill FIAN/TD/01-02 \ \ \ \ \ \

{}\hfill OHSTPY-HEP-T-01-004 \ \ \ \ \ \
 
{}\hfill hep-th/0103088  \ \ \ \ \ \

\vskip18cm

\section{Massive fields in $AdS_3$}

Let $\phi$ be arbitrary spin massive 
field propagating in $AdS_3$ space-time. 
Light-cone gauge action for this field can be cast into the following 
`nonrelativistic form'\cite{14}

\be\label{lcaction}
S_{l.c.}
= \int d^3 x  \partial^+\phi(-\partial^- +P^-)\phi\,,
\ee
\be\label{pm}
P^-=-\frac{\partial_z^2}{2\partial^+}
+\frac{1}{2z^2\partial^+}A\,,
\ee
where $P^-$ is the (minus) Hamiltonian\footnote{
We use parametrization of $AdS_3$ space in which
$ds^2=R^2(-dx_0^2+dx_1^2+dz^2)/z^2$, where $R$ is radius of $AdS$
geometry.
$\pm$ directions are defined as 
$x^\pm=(x^1 \pm x^0)/\sqrt{2}$ and 
$x^+$ is taken to be a light-cone time. We
adopt the conventions:
$\partial_z\equiv\partial/\partial z$,
$\partial^\pm=\partial_\mp \equiv \partial/\partial x^\mp$.}
while  $A$ 
is an operator acting only on spin indices of field $\phi$
and does not depending on space-time coordinates 
and  their derivatives.
We shall call the operator $A$ an $AdS$ mass 
operator\footnote{
$A$ is equal to zero only for 
those massless representations of AdS algebra $so(d-1,2)$ 
which can be realized as
irreducible representations for conformal algebra $so(d,2)$
\cite{24}.}.
By now this operator is known for (i) 
massive fields of arbitrary spin and arbitrary type of 
Young symmetry for $d>3$\cite{14} ;
(ii) massless fields of arbitrary spin in
totally (anti)symmetric
representations of $so(d-2)$ algebra\cite{14,15};
(iii) type IIB supergravity in $AdS_5 \times S^5$ and 
$AdS_3 \times S^3$ backgrounds\cite{16,12}.
So in order to fix light-cone action we should find 
$AdS$ mass operator.

Massive spin $s\geq 1$
field in $AdS_3$ has two physical degrees of freedom $\phi_1$,
$\phi_2$ which we combine into a 2-vector
\be\label{col}
\phi = \left(
\begin{array}{c}
\phi_1
\\
\phi_2
\end{array}
\right)\,.
\ee
The field $\phi$ can be labeled by spin $s$
and by either mass parameter $m$ or by lowest eigenvalue of
energy operator which we shall denote by $E_0$. 
We prefer to formulate 
our result in terms of $E_0$. 
So all that ones needs is to express $AdS$ mass operator in terms
of $E_0$ and $s$.
This operator is found to be
\be\label{aoper}
A= (E_0-s\sigma_3-1)^2 -\frac{1}{4}\,,
\ee
where $\sigma_3$ is the usual Pauli matrix. See Appendix for
some details.

As is well known an energy spectrum 
of spin $s\geq 1$ massive representation 
(see e.g. \cite{gunt}) corresponding to
{\it lowest weight unitary} representation of $so(2,2)$ algebra
is bounded from below
\be\label{bound}
E_0 \geq s+1\,.
\ee
As an illustration we would like to demonstrate how this bound
can be derived in framework of approach we discuss here. By 
product we will also explore how the discrete spectrum 
of energy operators is obtainable in the framework of
light-cone. To this end we use 
momentum basis of $p^+$ instead of coordinate basis of $x^-$
and make the following transformation of wave function
\be\label{tphi}
\phi = U \tilde{\phi}\,,
\qquad
U \equiv (p^+)^{\frac{1}{2}(z\partial_z -1)}\,,
\ee
The representation of
energy operator in $\tilde{\phi}$ takes then the form
\be\label{etot}
\hat{E} = \frac{1}{2}(H_1+H_2)\,,
\ee
where Hamiltonians $H_1$ and $H_2$ are given by
\be\label{ham1}
H_1 = -\frac{1}{2}\partial_z^2 +\frac{1}{2z^2}A +\frac{1}{2}z^2\,,
\ee
\be\label{ham2}
H_2 = -p^+ \partial_{p^+}^2 -\frac{1}{2}\partial_{p^+}
+\frac{1}{4p^+}(A-4B) + p^+\,,
\ee
and operator $B$ is given by (\ref{boper}). 
See Appendix for details. 
Introducing instead of momentum $p^+$ new variable $y$ by relation
$2p^+ = y^2$ we find the following representations for the Hamiltonians
\be
H_1 = -\frac{1}{2}\partial_z^2 
+\frac{1}{2z^2}(\kappa_1^2 - \frac{1}{4}) +\frac{1}{2}z^2\,,
\ee
\be
H_2 = -\frac{1}{2}\partial_y^2 
+\frac{1}{2y^2}(\kappa_2^2 - \frac{1}{4}) +\frac{1}{2}y^2\,,
\ee
where
\be\label{k1k2}
\kappa_1 \equiv E_0 -s -1\,,
\qquad
\kappa_2 \equiv E_0 + s -1\,,
\ee
and we restricted action of all operators
to $\tilde{\phi}_1$ (see (\ref{col}),(\ref{tphi})).
Spectra of the above Hamiltonians are well known
\be
E_1 = 2n_1 +|\kappa_1| +1\,,
\quad
E_2 = 2n_2 +|\kappa_2| +1\,,
\ee
$n_1,n_2= 0,1,\ldots $, 
so taking into account (\ref{etot}) we find an energy spectrum
\be\label{ensp}
E = n_1 + n_2 +\frac{|\kappa_1|+|\kappa_2|}{2}+1\,.
\ee
{}From formulas (\ref{k1k2}) we see that
the energy spectrum (\ref{ensp}), provided the inequality 
(\ref{bound}) holds true, takes the form 
\be
E = E_0 +n_1+n_2\,,
\ee
i.e. $E_0$ is indeed lowest eigenvalue of 
energy operator. The $\phi_2$ has the same
energy spectrum.
Note that helicity operator $\hat{J}$ being restricted to
$\tilde{\phi}_1$ and
$\tilde{\phi}_2$ takes the forms
\be\label{jtot}
\hat{J}= \frac{1}{2}(H_2 - H_1)\,,
\qquad
\hat{J}= \frac{1}{2}(H_1 - H_2)
\ee
respectively.
If (\ref{bound}) is satisfied the 
$\hat{J}$ has the following spectra

\be
J = s+ n_2 - n_1\,,
\qquad
J = -s+ n_1 - n_2
\ee
for $\phi_1$ and $\phi_2$ respectively.
So ground states of $\phi_1$ and $\phi_2$ have helicities
$+s$ and $-s$ respectively as it should be by construction.

Let us make comment on interrelation of light-cone fields $\phi_{1,2}$
with the ones of covariant approach. 
For simplicity consider spin $s=1$.
In covariant approach spin one 
massive field can be described by vector field $\phi^A$ and
Stueckelberg scalar field $\phi$\cite{ara}\footnote{
To our knowledge covariant description 
of massive spin $s>2$ fields in $AdS$ has not been 
worked out so far. Recent discussion of related issues may be found
in \cite{kl,buc}.}.
They satisfy well known massive spin one 
gauge invariant equations of motion in $AdS$ background 
which respect 
gauge transformations $\delta \phi^M = e^{\mu M}\partial_\mu \alpha$,
$\delta \phi =m\alpha$, 
where $e_\mu^M$ is a vielbein of $AdS$ background.
Upon imposing light-cone gauge only scalar
field $\phi$ and component of $\phi^M$ in $z$ direction, $\phi^z$,
survive. They are related with our light-cone fields (\ref{col}) as
follows $\phi_{1,2} = (\phi^z \pm \phi)/\sqrt{2}$. 
For the case of $s=1$ a comparison
of covariant and light-cone approaches gives relationship
between mass parameter $m$
and $E_0$: $E_0 = m+1$.
It seems highly likely that this relationship is valid for arbitrary spin
$s\geq 1$.

As a side remark let us point
out that all transformations of $AdS_3$ algebra do not mix field
$\phi_1$ with $\phi_2$ (\ref{col}).
Therefore we can make (anti)self dual projection either on $\phi_1$ 
or $\phi_2$ by simply taking one of these fields equal to zero. This gives
light-cone description of self dual 
arbitrary spin massive fields in $AdS_3$ space.
Discussion of such fields in $3d$ {\it flat} space at 
the level of covariant action may be found in \cite{25}.

\section{Compactification in $AdS$ spacetime}

In contrast to flat space
a compactification in $AdS$ spacetime cannot be achieved simply by
taking wave function to be periodic with respect to one of
{\it flat} coordinates. To illustrate 
what has just been said we consider equation of
motion of massive scalar field in $AdS_d$ and try to derive its
mass term via compactification from $AdS_{d+1}$.
Let us use Poincar\'e parametrization of $AdS_d$ 
\be\label{adsd}
ds_{AdS_d}^2  = \frac{1}{z^2}(dx^adx_a +dz^2)\,.
\ee
Here and below $x^a$, $a=0,1,\ldots d-2$, are
flat coordinates along boundary of $AdS_d$.
In this parametrization an action for field propagating in
$AdS_d$ takes a form

\be\label{2lcact}
S=\frac{1}{2}\int d^dx \phi\bigl(\Box_{d-1}
+\partial_z^2 -\frac{1}{z^2}A\bigr)\phi\,,
\ee
where from now on we use the notation
\be
\Box_{d-1} \equiv  - \partial_0^2 +\partial_1^2 
+\ldots +\partial_{d-2}^2\,.
\ee
This action gives the following equation of motion

\be\label{adseq}
(\Box_{d-1} +\partial_z^2 -\frac{1}{z^2}A)\phi(x,z) =0\,.
\ee
For the case of massive scalar field 
the AdS mass operator is expressible in terms of usual
mass parameter $m$ as follows

\be\label{AM}
A=(mR)^2+\frac{d(d-2)}{4}\,.
\ee
For the case of massless (conformal invariant) scalar field in 
$AdS_d$ one has $m^2=-\frac{d(d-2)}{4R^2}$ and this gives $A=0$.
Now we want to generate the mass-like term in (\ref{adseq}) 
starting with massless field in $AdS_{d+1}$ whose line element is 
taken to be

\be
ds_{AdS_{d+1}}^2  = \frac{1}{z^2}(dx^adx_a + dX^2 + dz^2)\,,
\ee
where $X$ is a coordinate of $AdS_{d+1}$ we are going to compactify.
Equation of motion of conformal invariant (massless) field is

\be
(\Box_{d-1} +\partial_z^2 +\partial_X^2)\phi(x,z,X)=0\,.
\ee
{}From this it is easy to see that if we take the wavefunction 
$\phi(x,z,X)$ to be periodic in $X$ and make Fourier expansion
then we get 
the following equation $(\Box - n^2)\phi_n(x,z)=0$ for $n$th mode. 
For $n =\!\!\!\!\! / \,0$ this equation however
does not describe massive field dynamics 
in $AdS_d$ as in the equation for massive field (see (\ref{adseq}))
the $AdS$ mass term $A=n^2$ 
should be accompanied by factor $z^{-2}$. 
This is the reason why naive compactifaction does not
work in $AdS$ spacetime. It turns out that in order to get
proper mass term one needs to include radial coordinate $z$ in a 
procedure of
compactifaction. To demonstrate this let us consider the simplest
case of compactifaction.

\subsection{
Compactification 
$AdS_{d+1}\rightarrow AdS_d\times S^1$. Scalar field
}

As before we start with Poincar\'e parametrization of
$AdS_{d+1}$

\be
ds_{AdS_{d+1}}^2  = \frac{1}{Z^2}(dx^adx_a + dX^2 + dZ^2)\,.
\ee
Now instead of radial coordinate $Z>0$ and coordinate $X$ which we want
to compactify we introduce new radial coordinate $z>0$ and angle coordinate
$\varphi$ on semi-circle $S^1$

\be
Z= z\sin\varphi\,,
\quad
X = z\cos \varphi\,,
\quad
0 \leq \varphi \leq \pi\,.
\ee
It is straightforward to see that line element of $AdS_{d+1}$ 
takes then the form

\be
ds_{AdS_{d+1}}^2  
= \frac{1}{\sin^2\varphi}(ds_{AdS_d}^2 +d\varphi^2)\,,
\ee
where line element of $AdS_d$ is as in (\ref{adsd})
i.e. $z$ is a usual radial coordinate of $AdS_d$.
Now we consider equation of motion for massless scalar field 
in $AdS_{d+1}$

\be\label{adseq2}
(\Box_{d-1} + \partial_X^2 + \partial_Z^2)\Phi(x,X,Z)=0\,.
\ee
Plugging Fourier expansion

\be
\Phi(x,X,Z) 
= \sum_{n \in \bf  Z} 
\frac{ e^{{\rm i}n\varphi} }{\sqrt{z}}\phi_n(x,z)
\ee
in (\ref{adseq2}) and using the textbook formula 

\be
\partial_X^2 +\partial_Z^2
=\frac{1}{z}\partial_z z\partial_z
+\frac{1}{z^2}\partial_\varphi^2
\ee
we get then the following equation for $n$th mode 

\be
(\Box_{d-1} + \partial_z^2 
-\frac{1}{z^2}(n^2 -\frac{1}{4}))\phi_n(x,z)=0\,.
\ee
This equation consists of properly normalized mass term (\ref{adseq})
and therefore
describes dynamics of field propagating in $AdS_d$. Comparison 
with (\ref{adseq}) gives the following value of $AdS$ mass operator
for $n$th mode $\phi_n$

\be\label{an}
A_n = n^2-\frac{1}{4}\,.
\ee
Taking into account the relation (\ref{AM}) we find a mass spectrum 
\be\label{massp}
m_n^2 = n^2 -\frac{(d-1)^2}{4}\,,
\ee
where the AdS radius $R$ is set equal to unity. Corresponding
lowest energy spectrum is

\be
E_{0 n} = n+\frac{d-1}{2}\,.
\ee
Note that 
since $n$ takes integer values  $A_n$ (\ref{an}) are never equal
to zero. This implies that upon compactification of $one$ dimension
fields of compactification $\phi_n$ do not 
satisfy conformal equations of motion in $AdS_d$.
Below we will demonstrate
that conformal mode survives only upon compactification of two 
spatial dimensions.

\subsection{
Compactification 
$AdS_{d+d'}\rightarrow AdS_d\times S^{d'}$. Scalar field}

Discussion of previous section can be generalized 
to spherical compactification 
in a rather straightforward way. In this section
we study mass spectra of massless and massive 
fields upon such a compactification.
To this end we start as before with 
Poincar\'e parametrization of line element of $AdS_{d+d'}$ 

\be
ds_{AdS_{d+d'}}^2  
= \frac{1}{Z^2}(dx^adx_a + dZ^2  + dX^\alpha dX^\alpha),
\ee
where
$\alpha=1,\ldots, d'$,
and $d$ is dimension of $AdS_d$ on which we make reduction
of field dynamics
while $d'$ is a number of coordinates we are going to compactify.
Instead of $Z$ and $X^\alpha$ we introduce new coordinates by relations

\be\label{spcor}
Z= z\sin\varphi,
\quad
X^\alpha = z u^\alpha \cos \varphi,
\quad
0 \leq \varphi \leq \pi,
\ee
where $u^\alpha$ is a unit vector $u^\alpha u^\alpha =1$.
In terms of the new coordinates $z$, $\varphi$,
$u^\alpha$ the line element of $AdS_{d+d'}$ 
takes the form 

\be
ds_{AdS_{d+d'}}^2  
= \frac{1}{\sin^2\varphi}(ds_{AdS_d}^2 +ds_{S^{d'}}^2)\,,
\ee
where line element of $AdS_d$ is as in (\ref{adsd}) while line element
of $S^{d'}$ takes the form 

\be
ds_{S^{d'}}^2 =d\varphi^2 +\cos^2\varphi (du^\alpha)^2\,.
\ee
As in previous case $z$ becomes radial coordinate
of $AdS_d$.
Now let us consider mass spectra of fields obtainable upon this
compactification. To this end we start with  equation of motion
in $AdS_{d+d'}$

\be\label{adseq4}
\Bigl(\Box_{d-1} + \partial_{X^\alpha}^2 + \partial_Z^2
-\frac{1}{Z^2}\hat{A}\Bigr)\hat{\Phi}(x,X,Z)=0\,,
\ee
where $\hat{\Phi}$ is a field propagating $AdS_{d+d'}$ 
while $\hat{A}$ is its $AdS$ mass operator.
Taking into account the textbook formula 

\be
\partial_{X^\alpha}^2 + \partial_Z^2 
=\frac{1}{z^{d'}}\partial_z z^{d'}\partial_z
+\frac{1}{z^2}\Delta_{S^{d'}}\,,
\ee
where $\Delta_{S^{d'}}$ is Laplace operator on $S^{d'}$ we find
that in terms of new field defined by
\be
\hat{\Phi}(x,X,Z) = z^{-d'/2}\Phi(x,z,u,\varphi)
\ee
the equation (\ref{adseq4}) takes desired form

\be
(\Box_{d-1} + \partial_z^2
-\frac{1}{z^2}A)\Phi(x,z,u,\varphi)=0\,,
\ee
where  $A$ is $AdS$ mass operator of 
the new field
$\Phi(x,z,u,\varphi)$. This field describes collection
of fields in $AdS_d$ which are obtainable by expanding in 
$u^\alpha$ and $\varphi$.
{}From the expressions above it is straightforward 
to see that $A$ is expressible in terms of $\hat{A}$ as follows

\be
A = \frac{1}{\sin^2\varphi}\hat{A} 
-\Delta_{S^{d'}} +\frac{(d'-1)^2}{4} -\frac{1}{4}\,.
\ee
Let us first consider the case of
massless (conformal invariant) field in $AdS_{d+d'}$, i.e.

\be
\hat{A}=0\,.
\ee
Decomposing the field 
$\Phi(x,z,u,\varphi)$ into harmonics of Laplace operator
$Y_l(u,\varphi)$

\be
\Phi(x,z,u,\varphi) = \sum_{l=0}^\infty
Y_l(u,\varphi)\phi_l(x,z)
\ee
and taking into account
spectrum of Laplace operator $\Delta_{S^{d'}}$ 

\be
\langle \Delta_{S^{d'}}\rangle  = - l(l+ d'-1)
\ee
we get then the following  value of $AdS$ mass operator for
mode $\phi_l(x,z)$
\be\label{Al}
A_l = (l+\frac{d'-1}{2})^2 -\frac{1}{4}\,.
\ee
We are able now to learn when 
conformal invariant mode  survives compactification.
{}From (\ref{Al}) 
we see that $A_l=0$ iff $d'=2$, $l=0$. Thus we conclude that
upon compactification 
$AdS_{d+d'}\rightarrow AdS_d\times S^{d'}$
conformal massless modes survives if and only if
we compactify two coordinates.

Now let us consider the case of massive field, i.e. when $\hat{A}$ 
takes the form given in (\ref{AM}). To find mass spectrum one needs
to use concrete form of Laplace operator $\Delta_{S^{d'}}$ in 
coordinates $u^\alpha$, $\varphi$
\be
\Delta_{S^{d'}}
=\frac{1}{\cos^2\varphi}\Delta_{S^{d'-1}} +\partial_\varphi^2
-(d'-1)\hbox{tg}\varphi\partial_\varphi\,,
\ee
where $\Delta_{S^{d'-1}}$ is a Laplace operator on unit 
$S^{d'-1}$ sphere defined by
$u^\alpha u^\alpha=1$. Making a rescaling
\be
\Phi(x,z,u,\varphi)
=(\cos \varphi)^{-(d'-1)/2} \phi(x,z,u,\varphi)
\ee
we find the following representation for $A$ in $\phi$
\begin{eqnarray}
A &=& \frac{1}{\sin^2\varphi}\hat{A} - \frac{1}{4}
\\
&-& \partial_\varphi^2 
+\frac{1}{\cos^2\varphi}\Bigl(
-\Delta_{S^{d'-1}}+\frac{(d'-2)^2-1}{4}\Bigr)\,.
\nonumber
\end{eqnarray}
Now decomposing the field $\phi(x,z,u,\varphi)$ into harmonics of
Laplace operator $Y_l(u)$ 

\be
\phi(x,z,u,\varphi) =\sum_{l=0}^\infty Y_l(u)\phi_l(x,z,\varphi)
\ee
we find the following $AdS$ mass operator for $\phi_l(x,z,\varphi)$

\be\label{Apt}
A_l =- \partial_\varphi^2
+\frac{\kappa^2-\frac{1}{4}}{\sin^2\varphi}
+\frac{(l+\frac{d'-2}{2})^2-\frac{1}{4}}{\cos^2\varphi}
-\frac{1}{4}\,,
\ee
where we have used the following representation for $\hat{A}$
(cf. (\ref{AM}))
\be
\hat{A} = \kappa^2 -\frac{1}{4}\,,
\qquad
\kappa \equiv \sqrt{m^2 +\frac{(d+d'-1)^2}{4}}\,.
\ee
The operator $A$ given by (\ref{Apt}) is nothing but Hamiltonian 
for P\"oschl-Teller potential and its spectrum is given by
\be
A_{L,l} = (2L+ l  + \kappa + \frac{d'}{2})^2 -\frac{1}{4}\,,
\quad
L=0,1,\ldots
\ee
Note that this formula is valid 
for values of $\kappa$, $d'$ and $l$ which satisfy
the inequalities
\be
\kappa >\frac{1}{2}\,,
\qquad
l+ \frac{d'-2}{2}>\frac{1}{2}\,.
\ee
Using formula (\ref{AM}) we get then corresponding
mass spectrum.

The above analysis can be generalized to arbitrary spin $s>0$ field in a 
straightforward way. For instance arbitrary
spin field in $AdS_4$ taken in light-cone gauge
satisfies an equation like (\ref{adseq2}).
Therefore in this case one gets a mass spectrum like (\ref{massp}).
For the case of spin field one needs however to analyse some details of
$AdS$ transformations. This will be done elsewhere. 
As a side remark let us point out that
mass spectra of compactification in $AdS$ spacetime
can also be studied in the framework of oscillator 
method \cite{gunt,gunwar}.

\section*{Appendix}

In this Appendix we present some details of deriving 
$AdS$ mass operator given in (\ref{aoper}). 
Representation for AdS algebra generators 
on the space of arbitrary spin fields taken
in light-cone gauge and propagating in $AdS_d$
was derived in \cite{14} where the light-cone form of
AdS field dynamics was discovered. $AdS_3$
algebra $so(2,2)$ taken in conformal algebra notation consists
of generators $P^\pm$, $K^\pm$, $J^{+-}$, $D$. 
They have the following representation\footnote{Normalization
of commutators of $so(2,2)$ algebra we use is given by
formula (B.10) in \cite{12}.}
\begin{eqnarray}
\label{3spp}
&&
P^+=\partial^+\,,
\\
&&
D = x^+ P^-+x^-\partial^+ + z\partial_z+\frac{1}{2}\,,
\\
\label{3sjpm}
&&
J^{+-}=x^+P^--x^-\partial^+\,,
\\
&&
K^+=-\frac{1}{2}(2x^+x^- + z^2)\partial^++x^+ D\,,
\end{eqnarray}
\be\label{km}
K^-=-\frac{1}{2}(2x^+ x^- + z^2) P^- +x^- D
+ \frac{1}{\partial^+}B
\ee
where $P^-$ is given by (\ref{pm}) and
operator $B$ acts only on spin degrees of freedom,
i.e. it does not depend on spacetime coordinates and their 
derivatives. As was mentioned above each massive 
representation 
can be labeled by $E_0$ and $s$.
So, all that one needs is to express the operators $A$ and $B$
in terms of $E_0$ and $s$.
This can easily be done by using technique of Casimir operators.
In conformal algebra notation 
Casimir operators of $so(2,2)$ algebra take the form
\be
Q_1 = 2\{P^-,K^+\} - (D+J^{+-})^2 
\ee
\be
Q_2 = 2\{P^+,K^-\}  - (D - J^{+-})^2\,.
\ee
Plugging in these expressions representation for generators
given in (\ref{pm}),(\ref{3spp})-(\ref{km}) we find 
\be\label{q1}
Q_1 = -A + \frac{3}{4}\,,
\qquad
Q_2 = -A + 4B + \frac{3}{4}\,.
\ee
On the other hand the values of these operators for
representations of $so(2,2)$ algebra
labeled by $E_0$ and $\pm s$ are given by
\begin{eqnarray}
\label{q1e}
&&
\langle Q_1 \rangle = -(E_0 - s\sigma_3)(E_0 - s\sigma_3 -2)\,,
\\
\label{q2e}
&&
\langle Q_2 \rangle = -(E_0 + s\sigma_3)(E_0 + s\sigma_3 -2)\,.
\end{eqnarray}
where we collected representations corresponding to helicites
$+s$ and $-s$
into 2-vector in the same way as in (\ref{col}).
Comparison of these expressions with (\ref{q1})
gives $AdS$ mass operator (\ref{aoper}) and operator $B$
\be\label{boper}
B = (1-E_0)s\sigma_3\,.
\ee
This completes light-cone description of massive field
in $AdS_3$. 

Note that to derive (\ref{q1e}),(\ref{q2e})
we re-express generators in light-cone basis in terms of ones
in energy basis
\begin{eqnarray}
&&\label{ppap}
P^+ = \frac{{\rm i}}{2}(
J^{w\bar{w}}  + J^{y\bar{y}} + J^{w\bar{y}} -  J^{\bar{w}y})\,,
\\
&&
P^- = \frac{{\rm i}}{2}( 
- J^{w\bar{w}} + J^{y\bar{y}} - J^{wy} + J^{\bar{w} \bar{y}})\,,
\\
&&
K^+ = \frac{{\rm i}}{2}( 
- J^{w\bar{w}} + J^{y\bar{y}} + J^{wy} -  J^{\bar{w}\bar{y}})\,,
\\
&&\label{kmap}
K^- = \frac{{\rm i}}{2}(  
J^{w\bar{w}} + J^{y\bar{y}} - J^{w\bar{y}} + J^{\bar{w}y})\,,
\\
&&
D = \frac{1}{2}(
-J^{w\bar{y}} - J^{wy} - J^{\bar{w}\bar{y}} - J^{\bar{w}y})\,,
\\
&&
J^{+-} = \frac{1}{2}(
J^{w\bar{y}} - J^{wy} - J^{\bar{w}\bar{y}} + J^{\bar{w}y})\,.
\end{eqnarray}
The $\hat{E}\equiv  J^{w\bar{w}}$ is nothing but energy operator,
$\hat{J} \equiv  J^{y\bar{y}}$ is a helicity operator, 
$J^{wy}$, $J^{w\bar{y}}$ are helicity de-boost operators
while
$J^{\bar{w}y}$, $J^{\bar{w}\bar{y}}$ are helicity boost operators.
In energy basis Casimir operators take the form

\be
Q_1 = -(J^{w\bar{w}} - J^{y\bar{y}})^2 
-2\{ J^{wy},J^{\bar{w}\bar{y}}\}\,,
\ee
\be
Q_2 = -(J^{w\bar{w}} + J^{y\bar{y}})^2 
-2\{ J^{w\bar{y}},J^{\bar{w}y}\}\,.
\ee
Using then a definition of ground state 
\be
\hat{E}|\hbox{vac}\rangle = E_0 |\hbox{vac}\rangle\,,
\quad
\hat{J}|\hbox{vac}\rangle = s\sigma_3  |\hbox{vac}\rangle\,,
\ee
\be
J^{wy}|\hbox{vac}\rangle = 0\,,
\qquad
J^{w\bar{y}}|\hbox{vac}\rangle = 0\,,
\ee
where the ground state 
$|\hbox{vac}\rangle$ is a 2-vector (see (\ref{col})),
we find eigenvalues of $Q_1$ and $Q_2$ (\ref{q1e}),(\ref{q2e}). 
Energy basis commutators of $so(2,2)$ we use are
\be
[J^{AB},J^{CE}] = \eta^{BC}J^{AE} +3\hbox{ terms}\,,
\ee
with nonvanishing $\eta^{y\bar{y}} = -\eta^{w\bar{w}}=1$.

In order to derive representation for $\hat{E}$ (\ref{etot}) 
we use relations (\ref{ppap})-(\ref{kmap}) to re-express $\hat{E}$ 
in terms of $P^\pm$ and $K^\pm$ given in 
(\ref{pm}),(\ref{3spp})-(\ref{km}),
restrict all expressions to surface of initial data 
$x^+=0$ and use momentum representation in which
$x^- = {\rm i}\partial_{p^+}$, $\partial^+ = {\rm i}p^+$.
Doing this we find 
$\hat{E} = (H_1 + H_2)/2$ where
\be
H_1 = -\frac{1}{2p^+}\partial_z^2 +\frac{1}{2p^+z^2}A +\frac{1}{2}p^+z^2\,,
\ee
\be
H_2= -\frac{z^2}{4p^+}(\partial_z^2 -\frac{1}{z^2}A) + p^+
\ee
$$
+\partial_{p^+}(-\partial_{p^+}p^+ +z\partial_z +\frac{1}{2}) 
-\frac{1}{p^+}B\,. \ \ \ \ \
$$
In basis of  $\tilde{\phi}$ (\ref{tphi})
these $H_1$ and $H_2$ take the form given
in (\ref{ham1}),(\ref{ham2}). 
In a similar way one can obtain the representation for 
$\hat{J}$ given in (\ref{jtot}).


\begin{thebibliography}{9}

\bibitem{1}
R.R. Metsaev and A.A. Tseytlin,
Phys.\ Rev.\ D {\bf 63} (2001) 046002
[hep-th/ 0007036].

\bibitem{2}
R.R. Metsaev, C.B. Thorn and A.A. Tseyt\-lin,
Nucl.\ Phys.\ B {\bf 596} (2001) 151
[hep-th/\-00\-09\-171].

\bibitem{3}
A.A. Tseytlin,
hep-th/ 0009226.

\bibitem{4}
R.~Kallosh,
hep-th/ 9807206.

\bibitem{5}
I.~Pesando,
JHEP{\bf 9811} (1998) 002
[hep-th/ 9808020].
Phys.\ Lett.\ B {\bf 485} (2000) 246
[hep-th/ 9912284].

\bibitem{6}
R. Kallosh and J. Rahmfeld,
Phys.\ Lett.\ B {\bf 443} (1998) 143
[hep-th/ 9808038].

\bibitem{7}
R. Kallosh and A.A. Tseytlin,
JHEP{\bf 9810} (1998) 016
[hep-th/ 9808088].

\bibitem{8}
R.R. Metsaev and A.A. Tseytlin,
Nucl.\ Phys.\ B {\bf 533} (1998) 109
[hep-th/ 9805028].
Phys.\ Lett.\ B {\bf 436} (1998) 281
[hep-th/9806095].

\bibitem{9}
R.~Kallosh, J.~Rahmfeld and A.~Rajaraman,
JHEP{\bf 9809} (1998) 002
[hep-th/9805217].

\bibitem{10}
R.~Roiban and W.~Siegel,
JHEP{\bf 0011} (2000) 024
[hep-th/0010104].

\bibitem{11}
R.R. Metsaev,
hep-th/0012026.


\bibitem{101}
P.~Claus, M.~Gunaydin, R.~Kallosh, J.~Rahmfeld and Y.~Zunger,
JHEP{\bf 9905} (1999) 019
[hep-th/9905112].


\bibitem{102}
N.~Berkovits,
hep-th/0008145.

\bibitem{103}
I.~Bandos, J.~Lukierski, C.~Preitschopf and D.~Sorokin,
Phys.\ Rev.\ D {\bf 61} (2000) 065009
[hep-th/9907113].


\bibitem{12}
R.R. Metsaev and A.A. Tseytlin,
hep-th/ 0011191.

\bibitem{13}
M.~A.~Vasiliev,
Phys.\ Lett.\ B {\bf 243} (1990) 378.

\bibitem{lop}
V.~E.~Lopatin and M.~A.~Vasiliev,
Mod.\ Phys.\ Lett.\ A {\bf 3} (1988) 257.

\bibitem{14}
R.~R.~Metsaev,
Nucl.\ Phys.\ B {\bf 563} (1999) 295
[hep-th/9906217].

\bibitem{24}
R.~R.~Metsaev,
Mod.\ Phys.\ Lett.\ A {\bf 10} (1995) 1719.

\bibitem{15}
R.~R.~Metsaev,
hep-th/0011112.


\bibitem{16}
R.R. Metsaev,
Phys.\ Lett.\ B {\bf 468} (1999) 65
[hep-th/9908114].


\bibitem{gunt}
M.~Gunaydin, G.~Sierra and P.~K.~Townsend,
Nucl.\ Phys.\ B {\bf 274} (1986) 429.

\bibitem{ara}
C.~Aragone, S.~Deser and Z.~Yang,
Annals Phys.\ {\bf 179} (1987) 76.

\bibitem{kl}
S.~M.~Klishevich,
hep-th/0002024.

\bibitem{buc}
I.~L.~Buchbinder and V.~D.~Pershin,
hep-th/0009026.



\bibitem{25}
I.~V.~Tyutin and M.~A.~Vasiliev,
Theor.\ Math.\ Phys.\ {\bf 113} (1997) 1244
[hep-th/9704132].

\bibitem{gunwar}
M.~Gunaydin, P.~van Nieuwenhuizen and N.~P.~Warner,
Nucl.\ Phys.\ B {\bf 255} (1985) 63.


\end{thebibliography}
\end{document}